\documentclass[aps,prl,reprint,showpacs]{revtex4-1}

\usepackage{graphics}
\usepackage{graphicx}
\usepackage{dcolumn}
\usepackage{bm}
\usepackage{color}
\usepackage{amssymb,amsmath}

\begin{document}

\title{Understanding the dynamics of photoionization-induced solitons in gas-filled hollow-core photonic crystal fibers}
\author{Mohammed F. Saleh}
\author{Fabio Biancalana}
\affiliation{Max Planck Institute for the Science of Light, G\"{u}nther-Scharowsky str. 1, 91058 Erlangen, Germany}

\begin{abstract}
We present in detail our developed model [Saleh \textit{et al.}, Phys. Rev. Lett. \textbf{107}] that governs pulse propagation in hollow-core photonic crystal fibers filled by an ionizing gas. By using perturbative methods, we find that the photoionization process induces the opposite phenomenon of the well-known Raman self-frequency red-shift of solitons in solid-core glass fibers, as was recently experimentally demonstrated [H\"olzer \textit{et al.}, Phys. Rev. Lett. \textbf{107}]. This process is only limited by ionization losses, and leads to a constant acceleration of solitons in the time domain with a continuous blue-shift in the frequency domain. By applying the Gagnon-B\'{e}langer gauge transformation, multi-peak `inverted gravity-like' solitary waves are predicted. We also demonstrate that the pulse dynamics shows the ejection of solitons during propagation in such fibers, analogous to what happens in conventional solid-core fibers. Moreover, unconventional long-range non-local interactions between temporally distant solitons, unique of gas plasma systems, are predicted and studied. Finally, the effects of higher-order dispersion coefficients and the shock operator on the pulse dynamics are investigated, showing that the resonant radiation in the UV [Joly \textit{et al.}, Phys. Rev. Lett. \textbf{106}] can be improved via plasma formation.
\end{abstract}
\maketitle

\section{Introduction}
The invention of photonic crystal fibers (PCFs) has led to a revolution in the field of nonlinear fiber optics \cite{Russell03}. Hollow-core PCFs (HC-PCFs) with the so-called kagom\'{e}-lattice claddings have recently become a superior host for the investigation of light-matter interactions between intense ultrashort optical pulses and gaseous or liquid media \cite{Russell06}. These fibers are characterized by a broadband transmission range with low group velocity dispersion (GVD), and a high confinement of light in the core \cite{Travers11}. Many important phenomena have already been observed in HC-PCFs. For instance, a drastic reduction in the Raman threshold has been shown in H$_{2}  $-filled HC-PCFs \cite{Benabid02a}. HC-PCFs have also been used to generate solitary pulses by backward stimulated Raman scattering \cite{Abdolvand09} and to observe self-similarity in the evolution of transient stimulated Raman scattering \cite{Nazarkin10}. Important applications such as high-harmonic \cite{Heckl09} and efficient deep UV generation \cite{Joly11} have been recently demonstrated in HC-PCFs filled with noble gases, which do not suffer of the limitations introduced by the Raman effect.

The concept of soliton self-frequency blue-shift has been introduced and predicted in \cite{Serkin93}. In tapered solid-core photonic crystal fibers, soliton blue-shift has been observed due to the variation of the zero-dispersion wavelength (ZDW) along the fiber \cite{Stark11}. In the photoionization regime, a limited (of only a few nanometers) ionization-induced blue-shift of guided ultrashort pulses has been predicted in conventional gas-filled photonic bandgap HC-PCFs, which suffer of large group velocity dispersion variations near the band edges, thus preventing the clear formation of solitary waves \cite{Serebryannikov07,Fedotov07}. Very recently, however, a sequence of emitted strong blue-shifted pulses has been observed in a groundbreaking experiment involving Ar-filled kagom\'{e}-style HC-PCFs, by using few $ \mu $J fs laser pulses \cite{Hoelzer11a,Hoelzer11b}. The use of the kagom\'{e} HC-PCF in these experiments was essential due to its unique guiding features. Such fibers possess an unusual broadband guidance, and a remarkably small group velocity dispersion (GVD) ($ |\beta_{2}|<10 $ fs$^{2}$/cm $ \equiv 1 $ ps$^{2}$/km  from 400 to 1000 nm) \cite{Nold10,Joly11,Hoelzer11b} in comparison to the traditional solid-core fibers. Moreover, the gas and waveguide contributions to the GVD can be balanced in the optical wavelength regime, unlike large-bore capillary-based systems, where the normal dispersion of the gas always dominates over the anomalous dispersion of the waveguide \cite{Nold10}.

Prior to our present work on the optical nonlinearities induced by photoionization, Geissler \textit{et. al.} modeled the photoionization process in terms of the full electric field of the pulse \cite{Geissler99}. Recently, we have presented a new model based on the evolution of the complex pulse envelope \cite{Saleh11}, which allowed several interesting predictions and it is more suitable for analytical and numerical investigation than other previously proposed models. The validity of the model has been carefully verified by using a numerical model \cite{Chang11} based on the unidirectional wave equation \cite{Kinsler10}. Our model allows us to use well-known analytical and numerical methods, developed to study pulse propagation in optical fibers, such as perturbation theory, the split-step Fourier method \cite{Agrawal07}, and soliton particle-like analogy \cite{Kosevich90,Mitchell96}. By using our new model: (i) we derive analytical expressions for the soliton self-frequency blue-shift in the presence and absence of the photoionization threshold of the gas, incidentally introducing the completely new concept of floating solitons; (ii) we predict the possibility of observing two-peak `inverted gravity-like' solitons; and finally (iii) we explore the effects of higher-order dispersion coefficients and shock term, demonstrating that a plasma-assisted blueshift can enhance the efficiency of resonant radiation in the UV, possibly improving previous experimental results \cite{Joly11}. Our theoretical results closely follow the experimental measurements in \cite{Hoelzer11b}.

The paper is organized as follows. In Sec. II, the governing equations of pulse propagation in HC-PCFs filled with an ionizing gas are reviewed. In Sec. III, the perturbation theory is applied to study the effect of photoionization on the soliton amplitude, the temporal location of the soliton peak, and the self-frequency shift. The possibility of observing two-peak `inverted gravity-like' solitons is discussed in Sec. IV. Section V is dedicated to study soliton dynamics in plasma. The effects of higher-order dispersion coefficients and shock term on pulse propagation are examined in Sec. VI. Our conclusions are provided in Sec. VII.

\section{Governing equations}
\begin{figure}
\includegraphics[width=8.6cm]{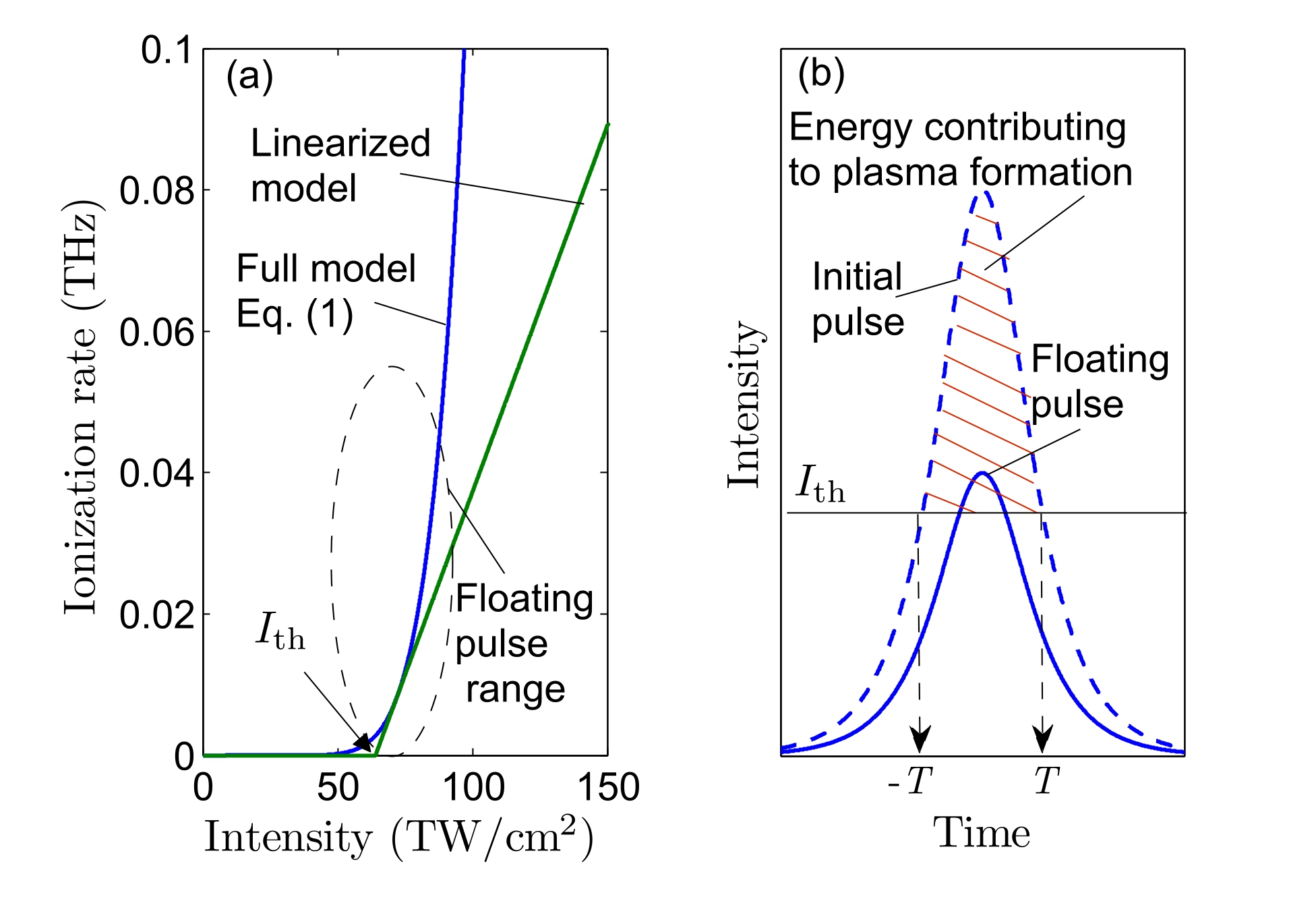}
\caption{(Color online). (a) Comparison of the dependence of the argon ionization rate on the pulse intensity using the full model of Eq. (\ref{eqW}) and the linearized model Eq. (\ref{eqL}). (b) A sketch that represents the attenuation of an initial pulse to a floating pulse above the threshold limit due to ionization-induced loss.  \label{fig0}}
\end{figure}

Photoionization can take place by either tunneling or multiphoton processes. These regimes are characterized by the Keldysh parameter $p_{\rm K}$ \cite{Wegener05,Sprangle02}. For optical pulses with intensities in the range $ 100 $ TW/cm$ ^{2} $, the Keldysh parameter is in the limiting regime between the tunneling or multiphoton processes. In this case, the Yudin-Ivanov modification of the Perelomov-Popov-Terent'ev (PPT) technique is considered to be the appropriate model since it accounts for both processes \cite{Yudin01,Popov04}. Experimental measurements \cite{Gibson90,Augst91} show that tunneling models \cite{Keldysh64,Ammosov86} can well describe photoionization processes in noble gases for optical-pulse intensities in the range of TW/cm$ ^{2} $. This is also confirmed by recent calculations \cite{Chang11} based on the tunneling Ammosov-Delone-Krainov (ADK) model \cite{Ammosov86} that reproduce simultaneous experimental results \cite{Hoelzer11b}. Hence, the time-averaged ionization rate $\mathcal{W}(I)$ can be expressed as  \cite{Keldysh64,Esarey97,Ammosov86}
\begin{equation}
\mathcal{W}(I)=c_{1}\,I ^{-1/4}\, \exp\left(  -c_{2}\,I^{-1/2}\right) ,\label{eqW}
\end{equation}
where $ c_{1} $ and $ c_{2} $ are constants, and $I$ is the laser pulse intensity. As shown in Fig. \ref{fig0}(a), Eq. (\ref{eqW}) predicts an ionization rate that has an exponential-like behavior for pulse intensities above a certain threshold value $ I_{\mathrm{th}} $. Ionization-induced loss that is due to the absorption of photons during plasma formation is \textit{proportional} to the ionization rate \cite{Sprangle02}. As a consequence, pulses with large intensities well above the threshold limit will have their intensities strongly driven back to near $ I_{\mathrm{th}} $, where the ionization-induced loss is drastically reduced. This allows us to \textit{linearize} Eq. (\ref{eqW}) by using the first-order Taylor series in the proximity of $ I_{\mathrm{th}} $, where pulses can survive without appreciable attenuation for relatively long time. This is our concept of floating pulses introduced in \cite{Saleh11} and shown in Fig. \ref{fig0}(b). Expanding Eq. (\ref{eqW}) in its linear regime results in
\begin{equation}
\mathcal{W}(I)\approx \tilde{\sigma}\,\left( I -I_{\mathrm{th}} \right)\, \Theta \left( I -I_{\mathrm{th}} \right),\label{eqL}
\end{equation}
where $ I_{\mathrm{th}} $ and $ \tilde{\sigma}$ can be related to $ c_{1} $, $ c_{2} $, and the expansion point, which is chosen to reproduce the physically observed threshold intensity of the considered gas \cite{Saleh11}. The purpose of the Heaviside function $ \Theta $ is to cut the ionization rate  below $ I_{\mathrm{th}} $. As shown in Fig. \ref{fig0}(a), the linearized model underestimates the ionization rate thence the ionization loss, in comparison to the full model. This yields surprisingly to a similar qualitative behavior between the two models even for $ I> I_{\mathrm{th}}$, since the ionization rate and the ionization loss are the key factors in the photoionization process.

By using Eq. (\ref{eqL}), one can show that propagation of light in a HC-PCF filled with an ionized Raman-active gas can be  modeled by the following coupled equations:
\begin{equation}
\left[ i\partial_{z}+\hat{D}(i\partial_{t})+\gamma_{\mathrm{K} } R(t)\otimes|\Psi(t)|^{2} -\dfrac{\omega_{\mathrm{p} }^{2}}{2k_{0}c^{2}}+i\tilde{\alpha}\right] \Psi  =0 \vspace{-3mm} , \label{eq1}
\end{equation}
\begin{equation}
\partial_{t}n_{\mathrm{e} }=[\tilde{\sigma}/A_{\mathrm{eff}}]\left[n_{\mathrm{T}}-n_{\mathrm{e}}\right] \left[|\Psi|^{2} -|\Psi|^{2}_{\mathrm{th}} \right] \Theta \left(|\Psi|^{2} -|\Psi|^{2}_{\mathrm{th}} \right)  , \label{eq2}
\end{equation}
where $\Psi(z,t)$ is the electric field \textit{envelope}, $z$ is the longitudinal coordinate along the fiber, $t$ is the time in a reference frame moves with the pulse group velocity, $\hat{D}(i\partial_{t})\equiv\sum_{m\geq 2}\beta_{m}(i\partial_{t})^{m}/m!$ is the full dispersion operator, $\beta_{m}$ is the $m$th order dispersion coefficient calculated at an arbitrary reference frequency $\omega_{0}$, $\gamma_{\mathrm{K}}$ is the  Kerr nonlinear coefficient of the gas, $R(t)= (1-\rho)\,\delta(t)+\rho\, h(t)$ is the normalized Kerr and Raman response function of the gas, $\delta(t)$ is the Dirac delta function, $\rho$ is the relative strength of the non-instantaneous Raman nonlinearity, $h(t)$ is the causal Raman response function of the gas \cite{Serebryannikov07,Agrawal07}, the symbol $\otimes$ denotes the time convolution $\left( A\otimes B\equiv\int A(t-t')B(t')dt'=\int B(t-t')A(t')dt'\right) $, $c$ is the speed of light, $k_{0}=\omega_{0}/c$, $ \omega_{0} $ is the input pulse central frequency, $\omega_{\mathrm{p} }=[e^{2} n_{\mathrm{e} }/(\epsilon_{0}m_{\mathrm{e} })]^{1/2}$ is the plasma frequency associated to an electron density $n_{\mathrm{e} }(t)$, $e$ and $m_{\mathrm{e}}$ are the electron charge and mass respectively, and $\epsilon_{0}$ is the vacuum permittivity, $ \tilde{\alpha}=\tilde{\alpha}_{1}+\tilde{\alpha}_{2} $ is the total loss coefficient, $ \tilde{\alpha}_{1} $ is the fiber loss, $ \tilde{\alpha}_{2}=\frac{A_{\mathrm{eff}}U_{I}}{2|\Psi|^{2}}\,\partial_{t}n_{\mathrm{e}} $ is the ionization-induced loss term, $A_{\mathrm{eff}} $ is the effective mode area, $ U_{I} $ is the ionization energy of the gas, $ |\Psi|^{2}=I A_{\mathrm{eff}} $, $ |\Psi|_{\mathrm{th}}^{2}=I_{\mathrm{th}} A_{\mathrm{eff}} $, and $n_{\mathrm{T} }$ is the total number density of ionizable atoms in the fiber, associated to the maximum plasma frequency $\omega_{\mathrm{T}}\equiv[e^{2} n_{\mathrm{T} }/(\epsilon_{0}m_{\mathrm{e}})]^{1/2}$. The recombination process is safely neglected in these coupled equations since the pulse duration (of the order of tens of fs) is always shorter than the recombination time \cite{Sprangle02,Wood93}. However, the recombination effects as well as the rearrangement of the electronic plasma due to ponderomotive forces should be included  in the case of long-pulse propagation, making the problem considerably difficult \cite{Bellan06,Dendy90}.

Introducing the following rescalings and redefinitions: $\xi\equiv z/z_{0}$, $\tau\equiv t/t_{0}$, $\Psi_{0}\equiv[\gamma_{\mathrm{K} }z_{0}]^{-1/2}$, $\psi\equiv \Psi/\Psi_{0}$, $|\psi_{\mathrm{th}}|\equiv |\Psi|_{\mathrm{th}}/\Psi_{0}$, $r(\tau)\equiv R(t)\,t_{0}$, $\phi\equiv \frac{1}{2}k_{0}z_{0}\,[\omega_{\mathrm{p}}/\omega_{0}]^{2}$, $\phi_{\mathrm{T}}\equiv \frac{1}{2}k_{0}z_{0}\,[\omega_{\mathrm{T}}/\omega_{0}]^{2}$, $\sigma\equiv\tilde{\sigma}\,t_{0}/[ A_{\mathrm{eff}}\gamma_{\mathrm{K}}\,z_{0}]$, and $ \kappa\equiv U_{I}\, \tilde{\sigma}\,\epsilon_{0}\,m_{\mathrm{e}}\, \omega_{0}^{2}/[k_{0}\,e^{2}]$, where $z_{0}\equiv t_{0}^{2}/|\beta_{2}(\omega_{0})|$ is the second-order dispersion length at the reference frequency $\omega_{0}$, and $t_{0}$ is the input pulse duration \cite{Agrawal07}. Hence, the two coupled equations for floating pulses can be replaced by
\begin{equation}
\left[ i\partial_{\xi}+\hat{D}(i\partial_{\tau})+r(\tau)\otimes|\psi(\tau)|^{2}-\phi+i\alpha\right]\psi=0 \vspace{-6mm} ,\label{eq3}
\end{equation}
\begin{equation}
\partial_{\tau}\phi=\sigma(\phi_{\mathrm{T}}-\phi) \left[|\psi|^{2} -|\psi|^{2}_{\mathrm{th}} \right] \Theta \left(|\psi|^{2} -|\psi|^{2}_{\mathrm{th}} \right),\label{eq4}
\end{equation}
where $ \alpha=\kappa\,\left(\phi_{\mathrm{T}}-\phi \right)\, \left[1-|\psi|^{2}_{\mathrm{th}}/|\psi|^{2} \right]\, \Theta \left(|\psi|^{2} -|\psi|^{2}_{\mathrm{th}}\right)   $, and the fiber losses are neglected.

\section{Perturbation theory for solitons}
The effect of the Raman and ionization perturbations on the soliton dynamics can be studied using the perturbation theory described in standard textbooks (e.g. \cite{Agrawal07}). Neglecting higher-order dispersion coefficients, i.e. $\beta_{m>2}=0  $, the solution of a perturbed nonlinear Schr\"{o}dinger equation, $i\partial_{\xi} \psi+\frac{1}{2}\,\partial^{2}_{\tau} \psi+|\psi|^{2} \psi=i\,\epsilon(\psi) $ is assumed to be $\psi(\xi,\tau)=A(\xi)\,{\mathrm{sech}}\left[A(\xi)(\tau-\tau_{\mathrm{p}}(\xi))\right]e^{-i\delta(\xi)\tau}$, where  $A(\xi)  $ is the soliton amplitude, $\tau_{\mathrm{p}}(\xi)$ is the temporal location of the soliton peak, $\delta(\xi)$ is self-frequency shift, and $\epsilon(\psi)$ is the perturbation function. Substituting this {\textit{Ansatz}} in Eqs. (\ref{eq3},\ref{eq4}), the following ordinary differential equations (ODEs) are obtained:
\begin{eqnarray}
&& \frac{\partial\delta}{\partial\xi}=-\mathrm{Im}\left\lbrace \int_{-\infty}^{+\infty}\epsilon(\psi)\tanh\left[A(\tau-\tau_{\mathrm{p}})\right]\,\psi^{*}\,d\tau\right\rbrace ,\label{ode1} \\
&& \frac{\partial\tau_{\mathrm{p}}}{d\xi}=-\delta+\frac{1}{A}\mathrm{Re}\left\lbrace \int_{-\infty}^{+\infty}\epsilon(\psi)(\tau-\tau_{\mathrm{p}})\,\psi^{*}\,\partial\tau\right\rbrace ,\label{ode2}\\
&&\frac{\partial A}{\partial\xi}=\mathrm{Re}\left\lbrace \int_{-\infty}^{+\infty}\epsilon(\psi)\,\psi^{*}\,d\tau\right\rbrace ,\label{ode3}
\end{eqnarray}
where $ \mathrm{Re} $ and $ \mathrm{Im} $ stand for the real and imaginary parts.

In order to extract useful analytical information from Eqs. (\ref{eq3},\ref{eq4}), we will start with the simple case where we assume that the ionization loss and the threshold are negligible. Ignoring ionization loss is unrealistic unless the pulse under consideration has its maximum amplitude just above the threshold, in which case it does not feel strong plasma-induced decay. Hence, Eq. (\ref{eq4}) can be solved analytically, $ \phi(\tau)=\phi_{\mathrm{T}}\left\lbrace 1 - \exp\left[-\sigma\int_{-\infty}^{\tau}|\psi(\tau')|^{2}d\tau'\right]\right\rbrace $, with an initial condition $\phi(-\infty)=0$, corresponding to the absence of any plasma before the pulse arrives.  For a small ionization cross-section, $ \phi(\tau)\simeq \eta\int_{-\infty}^{\tau}|\psi(\tau')|^{2}d\tau' $, where $\eta\equiv\sigma\phi_{\rm T}$. Moreover, in the long-pulse limit $|\psi(\tau-\tau')|^{2}\simeq |\psi(\tau)|^{2}-\tau'\partial_{\tau}|\psi(\tau)|^{2}$  \cite{Agrawal07}. This allows the two coupled equations to be reduced to a single partial integro-differential equation:
\begin{equation}
i\partial_{\xi}\psi+\hat{D}(i\partial_{\tau})\psi+|\psi|^{2}\psi-\tau_{\mathrm{R}}\psi\partial_{\tau}|\psi|^{2}-\eta\psi\int_{-\infty}^{\tau}\!\!\!\!\!\!\!\!\!|\psi|^{2}d\tau' =0,
\label{eq5}
\end{equation}
where $\tau_{\rm R}\equiv\int_{0}^{\infty} \tau'\,r(\tau')\, d\tau'$. This equation shows clearly that the effect of ionization is almost exactly opposite to that of the Raman effect: the fourth term in Eq. (\ref{eq5}) involves a {\em derivative} of the field intensity, while the fifth term involves an {\em integral} on the same quantity. Hence, one can conjecture that the ionization perturbation will lead to a soliton self-frequency blue-shift, instead of a redshift. However, this is not an exact analogy because of thermodynamical reasons. In fact, the Raman effect acts on a soliton by constantly decreasing its central frequency, leaving its pulse shape in the time domain and in the frequency domain undeformed - in other words the soliton continuously  converts its coherent energy into incoherent optical phonons in the medium, and therefore the number of photons contained in the soliton does not change, and the overall entropy increases according to the second law of thermodynamics. However, in the photoionization-induced soliton blueshift described above, the soliton coherently receives energy from the medium (the plasma in this case). For this to be physically possible without breaking the second law of thermodynamics, the number of photons in the soliton must necessarily decrease - thus the \textit{unavoidable} plasma-induced losses cannot be eliminated on principle. The stronger the initial pulse intensity, the stronger the soliton blueshift, and the larger the plasma-induced losses, so that Eq. (\ref{eq5}), which assumes small losses, would lose its validity. However, in the following we shall prove numerically that when solitons decrease their amplitudes to just above the ionization threshold, their losses are very limited thence they can propagate for relatively long distances. This led us to naturally introduce the concept of {\em floating solitons}, i.e. those solitary pulses with maximum amplitude just a little above the threshold--- the regime where ionization losses can be neglected and Eq. (\ref{eq5}) maintains its validity \cite{Saleh11}.

Solving Eq. (\ref{eq5}) with a perturbation function $\epsilon(\psi)=-i\psi\left[\tau_{\mathrm{R}}\partial_{\tau}|\psi|^{2}+\eta\int_{-\infty}^{\tau}|\psi(\tau')|^{2}d\tau'\right]$, results in $ A(\xi)=A(0)=A_{0} $, $\delta(\xi)=\delta_{\rm Raman}(\xi)+\delta_{\rm ion}(\xi)=-g\,\xi  $, $\tau_{\rm p}(\xi)=g\,\xi^{2}/2$, and $ g= g_{\rm red}+g_{\rm blue}$, where $g_{\rm red}=+(8/15)\tau_{\rm R}A_{0}^{4}$ and $g_{\rm blue}=-(2/3)\eta A_{0}^{2}$ \cite{Saleh11}. Note that $g$ can be positive, negative or even zero, depending on the value of $\eta$, $\tau_{\rm R}$ and $A_{0}$. The precise rate for the self-frequency blue-shift can be obtained using the exact formula of $\phi(\tau)$, $g'_{\rm blue}=\sigma^{-2}A_{0}^{-1}\phi_{\rm T}\left[(1-\sigma A_{0})-(1+\sigma A_{0})\exp(-2\sigma A_{0})\right]$, which tends to $g_{\rm blue}$ for small values of $\sigma$, but start to differ considerably from it for $A_{0}>\sigma^{-1}$.

A further step can be applied to take into account the effect of the threshold intensity on the blue-shift rate. For floating pulses, the ionization loss is not large and can be neglected to a good approximation. For such pulses, only a small portion of energy above the threshold intensity contributes to the electron density build-up. In this case, $ \phi(\tau)\simeq \eta\int_{-T}^{\tau'\leq T}\left[ |\psi(\tau')|^{2}-|\psi|^{2}_{\mathrm{th}}\right]d\tau' $, where $ T=A_{0}^{-1}  \mathrm{sech}^{-1} \left[|\psi|_{\mathrm{th}}/A_{0} \right] $. This formulation embeds the Heaviside function introduced in Eq. (\ref{eq4}). Replacing the fifth term in Eq. (\ref{eq5}) with the new definition of $ \phi(\tau) $ and solving Eq. (\ref{ode1}), results in
\begin{equation}
g_{\rm blue}=-\eta A_{0}^{2}\left[ \dfrac{2}{3}\,\mathrm{tanh}^{3}\,\theta+\dfrac{|\psi|^{2}_{\mathrm{th}}}{A_{0}^{2}}\left( \theta \,\mathrm{sech}^{2}\theta-\mathrm{tanh}\,\theta\right) \right],
\end{equation}
with $ \theta=A_{0}T $. It can be seen that this expression tends to $ -(2/3)\eta A_{0}^{2} $ for small $ |\psi|_{\mathrm{th}} $.

The threshold intensity effect on the blue-shift rate can also be alternatively estimated by writing Eq. (\ref{eq5}) as
\begin{equation}
i\partial_{\xi}\psi+\hat{D}(i\partial_{\tau})\psi+|\psi|^{2}\psi-\tau_{\mathrm{R}}\psi\partial_{\tau}|\psi|^{2}-\tilde{\eta}\psi\int_{-\infty}^{\tau}\!\!\!\!\!\!\!\!\!|\psi|^{2}d\tau' =0,
\label{eq5prime}
\end{equation}
where $ \tilde{\eta}=\eta\, \mu $ and
\begin{equation}
\mu=\frac{\int_{-T}^{T}\left[ |\psi(\tau')|^{2}-|\psi|^{2}_{\mathrm{th}}\right] d\tau'}{\int_{-\infty}^{\infty}|\psi(\tau')|^{2}d\tau'}.
\end{equation}
The factor $ \mu $ represents the ratio between the pulse energy contributing to plasma formation and the total energy of the pulse, see Fig. \ref{fig0}(b). Hence, it takes into account the overestimation of the plasma density by neglecting the threshold intensity. As a consequence, one can simply prove that $g_{\rm blue}= -(2/3)\tilde{\eta} A_{0}^{2} $.

\begin{figure}
\includegraphics[width=8.6cm]{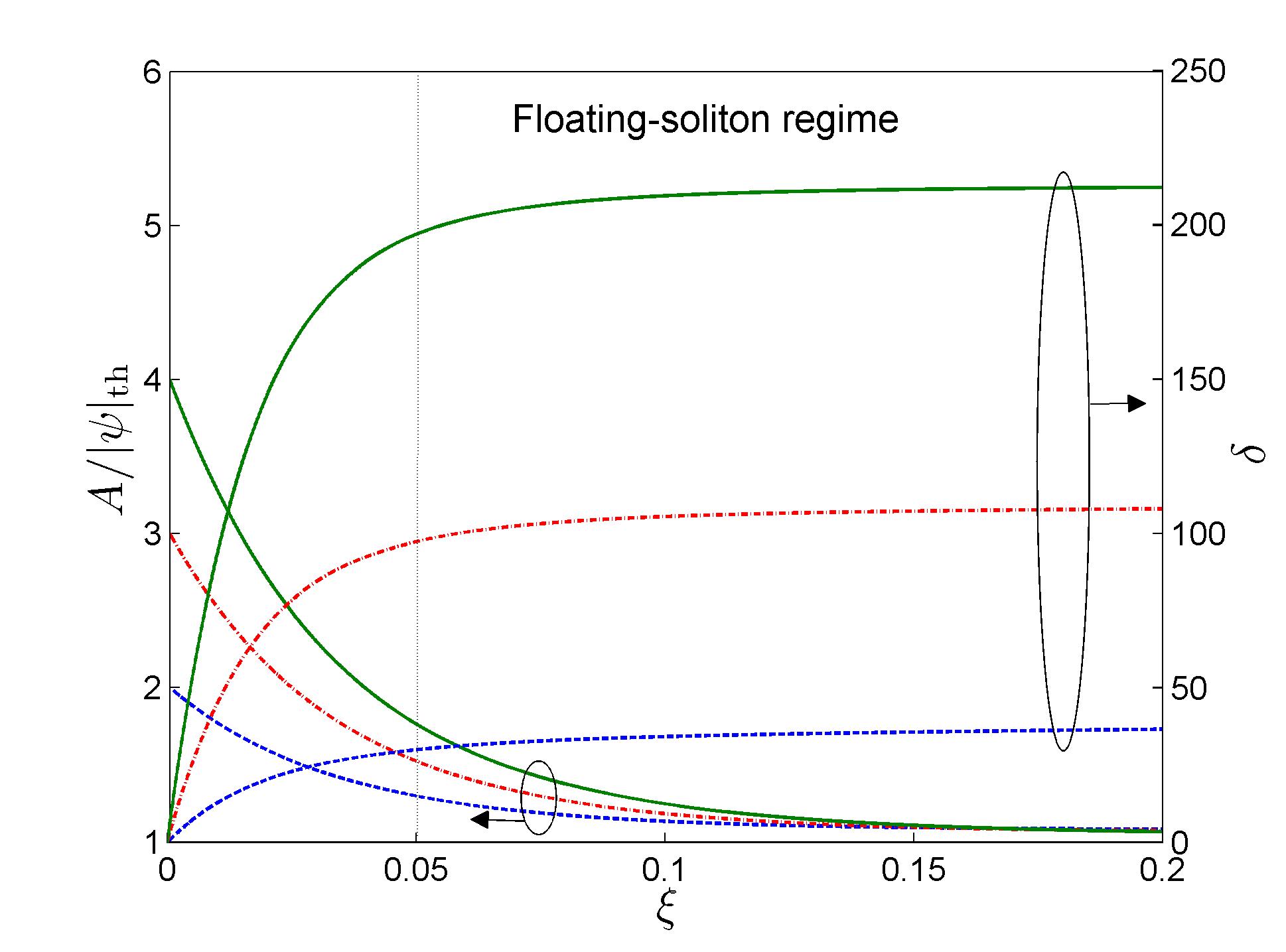}
\caption{(Color online). The spatial dependence of the soliton amplitude and frequency shift during a photoionization process for different initial pulse amplitudes.}\label{fig1}
\end{figure}

Moreover, the effect of the photoionization loss on the soliton amplitude and frequency shift can be studied numerically by solving the above ODEs with $\epsilon(\psi)=-i\psi\left[\eta\int_{-T}^{\tau'\leq T}\left( |\psi(\tau')|^{2}-|\psi|^{2}_{\mathrm{th}}\right)d\tau'-i\alpha\right]$, where $ T $ is determined via its previous definition using the initial pulse amplitude, and the variation of the position $ T $ due to losses is neglected,
\begin{equation}
\begin{array}{ll}
\dfrac{\partial A}{\partial\xi}= &-2\kappa\,\phi_{\mathrm{T}}\,\left(A\,\mathrm{tanh}\,\vartheta -|\psi|_{\mathrm{th}}^{2}\,T\right)\vspace{0.5mm}  \\
 \dfrac{\partial\delta}{\partial\xi}=& \eta\, A^{2}\left[ \dfrac{2}{3}\,\mathrm{tanh}^{3}\,\vartheta + \dfrac{|\psi|_{\mathrm{th}}^{2}}{A^{2}}\left(\vartheta \,\mathrm{sech}^{2}\vartheta-\mathrm{tanh}\,\vartheta\right) \right],
\end{array}
\end{equation}
$ \vartheta=AT $. These equations can be solved numerically to determine the spatial dependence of the soliton amplitude and frequency shift, as shown in Fig. \ref{fig1}. Pulses with initial large intensities $\left( A_{0}^{2} > |\psi|^{2}_{\mathrm{th}}\right)$ have a boosted self-frequency blue-shift. However, the ionization loss suppresses the soliton intensity after a short propagation distance to the floating-soliton regime, where the soliton can propagate for a long propagation distance with a limited blue-shift and negligible loss. The maximum frequency shift is achieved when the soliton intensity goes below the photoionization threshold.

\section{Inverted gravity-like bound states}
The previous conclusions on the strong analogies between photoionization shift and Raman shift of solitons allow us to construct a novel kind of solitary wave that can be supported by both the Kerr and the ionization nonlinearities. This solitary waves will be analogous to the Raman bound solitons discovered experimentally in \cite{Podlipensky07} and explained theoretically in \cite{Hause10a,Tran10,Hause10b}. In the non-inertial reference frame of an accelerated soliton (with a new time coordinate $\zeta\equiv\tau-g\,\xi^{2}/2$), and by using the so-called Gagnon-B\'{e}langer gauge transformation $\psi(\xi,\tau)=f(\zeta)\exp\left[i(q-g^{2}\,\xi^{2}/3+g\,\tau)\,\xi\right]$ \cite{Gagnon90}, Eq. (\ref{eq5prime}) can be written as an integro-differential equation,
\begin{equation}
\frac{1}{2}\partial_{\zeta}^{2}f-U(\zeta)\,f=q\,f, \label{gagnon}
\end{equation}
where ionization loss and higher-order dispersion are neglected, $ U(\zeta)= g\,\zeta -|f|^2+\tau_{R}\partial_{\zeta}|f|^{2} +\tilde{\eta}\int_{-\infty}^{\zeta}|f|^{2} d \zeta'$ is a potential \cite{Gorbach07a,Gorbach07b}, in which the first term is gravity-like, $q= A_{0}^{2}/2-\tilde{\eta} A_{0}$ is the soliton wavenumber, and $ A_{0} $ is the amplitude of the one-peak solitary solution of Eq. (\ref{gagnon}). In the case when only Raman effect was present, a special kind of nonlinear metastable bound states have been found as a general solution for Eq. (\ref{gagnon}) \cite{Akhmediev96,Gorbach07a,Gorbach07b}. Such bound states are multi-peak stationary states (in the non-inertial reference frame moving with acceleration $g_{\rm red}$), which are due to the Kerr effect complemented by the Raman nonlinearity \cite{Hause10a,Tran10,Hause10b}.

\begin{figure}
\includegraphics[width=8.6cm]{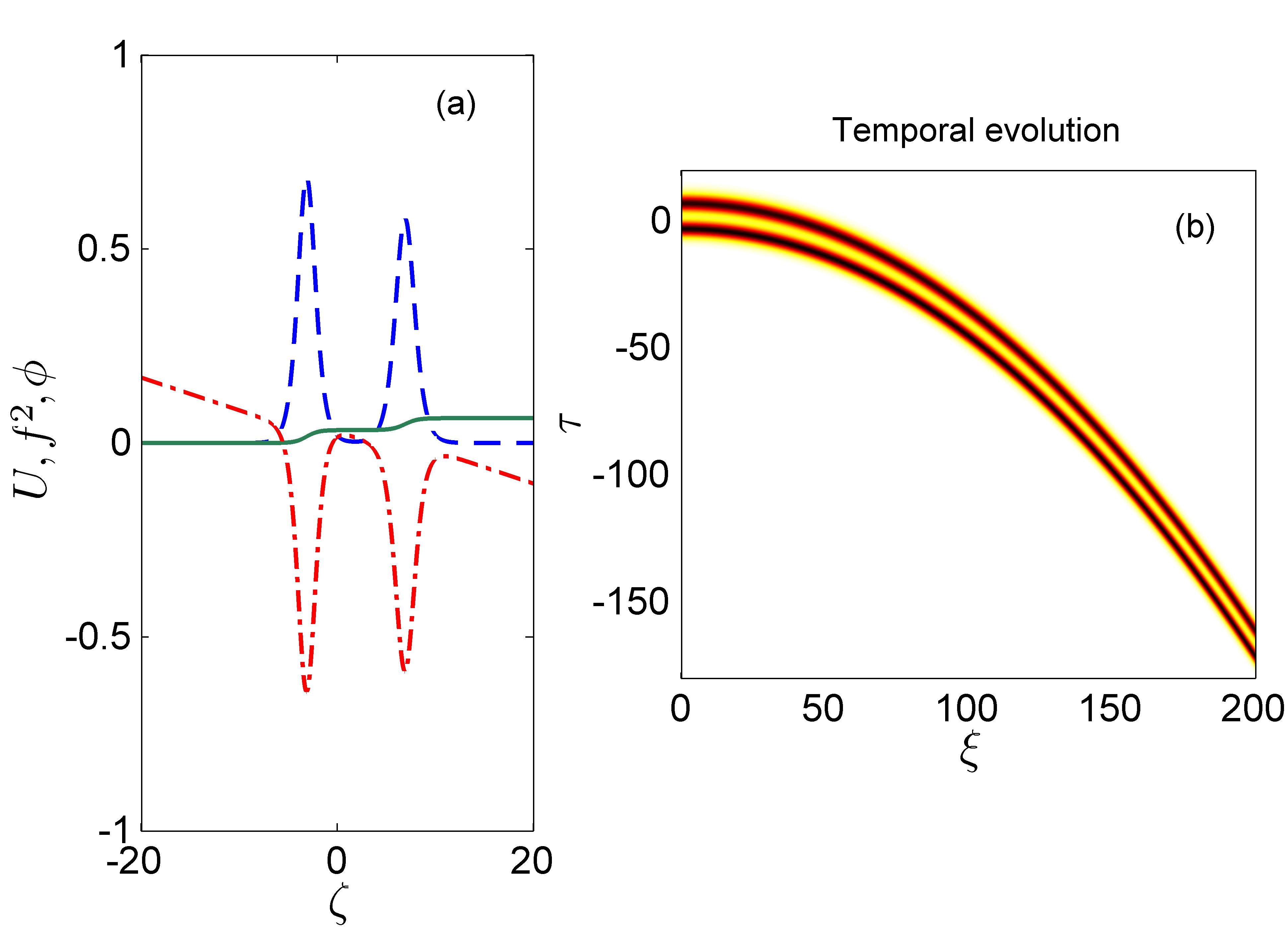}
\caption{(Color online). (a) A two-peak soliton stationary solution for a Raman-inactive gas with $ \tilde{\eta}=0.02 $ and $ q=0.3 $. The dashed-dotted red, dashed blue, and solid green curves represent the nonlinear potential $ U $, the pulse intensity $ f^{2} $, and the ionization field $ \phi $, respectively. (b) Temporal evolution of the two-peak soliton in a gas-filled HC-PCF fiber.
\label{fig2}}
\end{figure}

In the presence of a Raman-inactive gas (such as Argon) inside the HC-PCF, we have seen that the ionization process leads to a soliton acceleration in the time domain, and to a linear shift of its frequency toward the blue with a rate $g_{\rm blue}$. In this case, solitons will feel an `anti-gravity' field, and the solution of Eq. (\ref{gagnon}) can be a multi-peak stationary state with a negative slope opposite to the Raman case. An example of a two-peak bound state found numerically by using the shooting method is depicted in Fig. \ref{fig2}(a). Multi-peak solitary solutions analogous to those found in \cite{Tran10} can also be obtained. The propagation of the two-peak bound state found in Fig. \ref{fig2}(a) in a long HC-PCF is shown in Fig. \ref{fig2}(b).

The characteristics of a two-peak solitary solution --- the temporal separation $\zeta_{0}$ and the amplitude imbalance $R$ between the two peaks --- can be analytically determined. Consider a general two-peak solitary solution,
\begin{equation}
f(\zeta)=A_{1}\mathrm{sech}\left[A_{1} \left(\zeta-\zeta_{1} \right) \right] + A_{2}\mathrm{sech}\left[A_{2} \left(\zeta-\zeta_{2} \right) \right],\label{eq2peaksol}
\end{equation}
where $ A_{2}=R\,A_{1} $ and $\zeta_{0}=\zeta_{2}-\zeta_{1}$. In principle, a set of three algebraic equations is needed to determine $A_{1}$, $ \zeta_{0} $ and $ R $. By substituting the above solution in Eq. (\ref{gagnon}), the first equation is obtained,
\begin{equation}
-g_{\rm blue}\zeta_{0}=A_{1}^{2}(1-R^{2})/2+\tilde{\eta} A_{1}(1+R),\label{eqZeta0}
\end{equation}
where $ \tilde{\eta} $ is assumed to be the same for the two solitons. The last term on the right-hand side is due to the integration constant in the nonlinear potential $ U(\zeta) $, and it is not present in the analogous expression for the pure Raman-bound states [i.e. $-g_{\rm red}\zeta_{0}=A_{1}^{2}(1-R^{2})/2$] since the Raman effect is modeled in the equations by a derivative. Multiply Eq. (\ref{gagnon}) by $f^{*}$ and integrate its both sides, results in
\begin{equation}
\int_{-\infty}^{\infty}\left[ \frac{1}{2}f^{*}\partial_{\zeta}^{2}f-U(\zeta)\,|f|^{2} \right] d\zeta=q\int_{-\infty}^{\infty}|f|^{2} d\zeta, \label{eqInt}
\end{equation}
where the superscript $ * $ denotes the complex conjugate. A second equation can be attained by substituting Eq. (\ref{eq2peaksol}) into Eq. (\ref{eqInt}),
\begin{equation}
A_{1}^{2}(1+R^{3})/2-\tilde{\eta} A_{1}(1+2R+R^{2})-q(1+R)\approx 0. \label{eqR}
\end{equation}
The set of three algebraic equations is completed by Eq. (\ref{gagnon}) evaluated at $ \zeta=0 $. Hence, the variables $A_{1}$, $ \zeta_{0} $ and $ R $ can be determined for a certain value of $ q $.

\begin{figure}
\includegraphics[width=8.6cm]{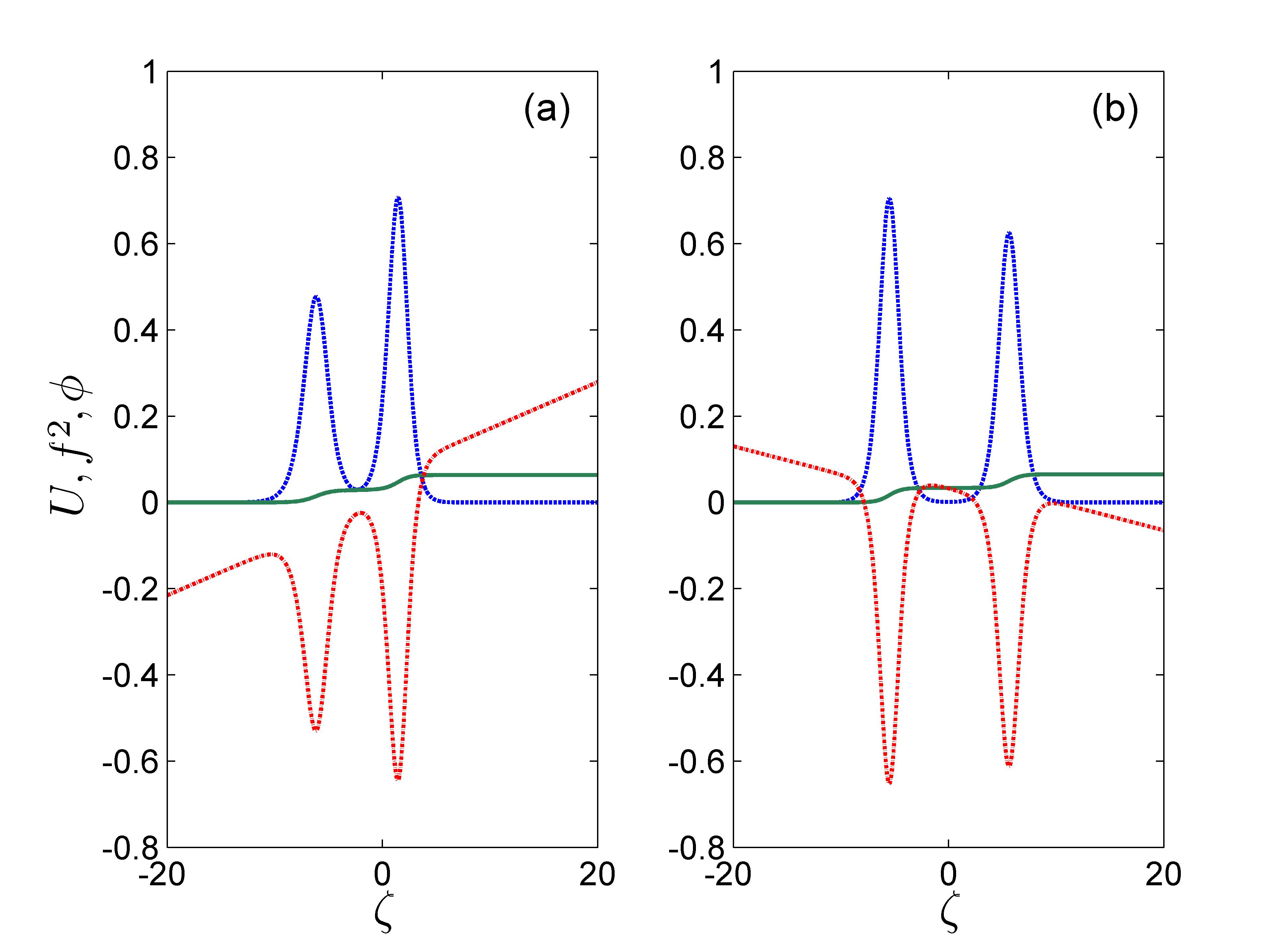}
\caption{(Color online). A two-peak soliton stationary solution in a Raman-active gas with $ \tilde{\eta}=0.02 $ and $ q=0.3 $. (a) Gravity-like bound states with $ \tau_{\mathrm{R}}=0.1 $. (b) Inverted gravity-like bound states with $ \tau_{\mathrm{R}}=0.01 $. The dashed-dotted red, dashed blue, and solid green curves represent the nonlinear potential $ U $, the pulse intensity $ f^{2} $, and the ionization field $ \phi $, respectively. \label{fig3}}
\end{figure}

Finally in Raman-active gases, solitons may feel either a `gravity-like' or an `inverted-gravity-like' potential based on the sign of the total gravity acceleration $ g= g_{\rm red}+g_{\rm blue}$. Hence, the two-peak solitary solution can have either positive or negative slopes as depicted in the panels (a,b) of Fig. \ref{fig3}, respectively. Two-peak solitons with positive (negative) slope are obtained when Raman (photoionization) process is dominant.

\section{Soliton dynamics in plasma}
\begin{figure}
\includegraphics[width=8.6cm]{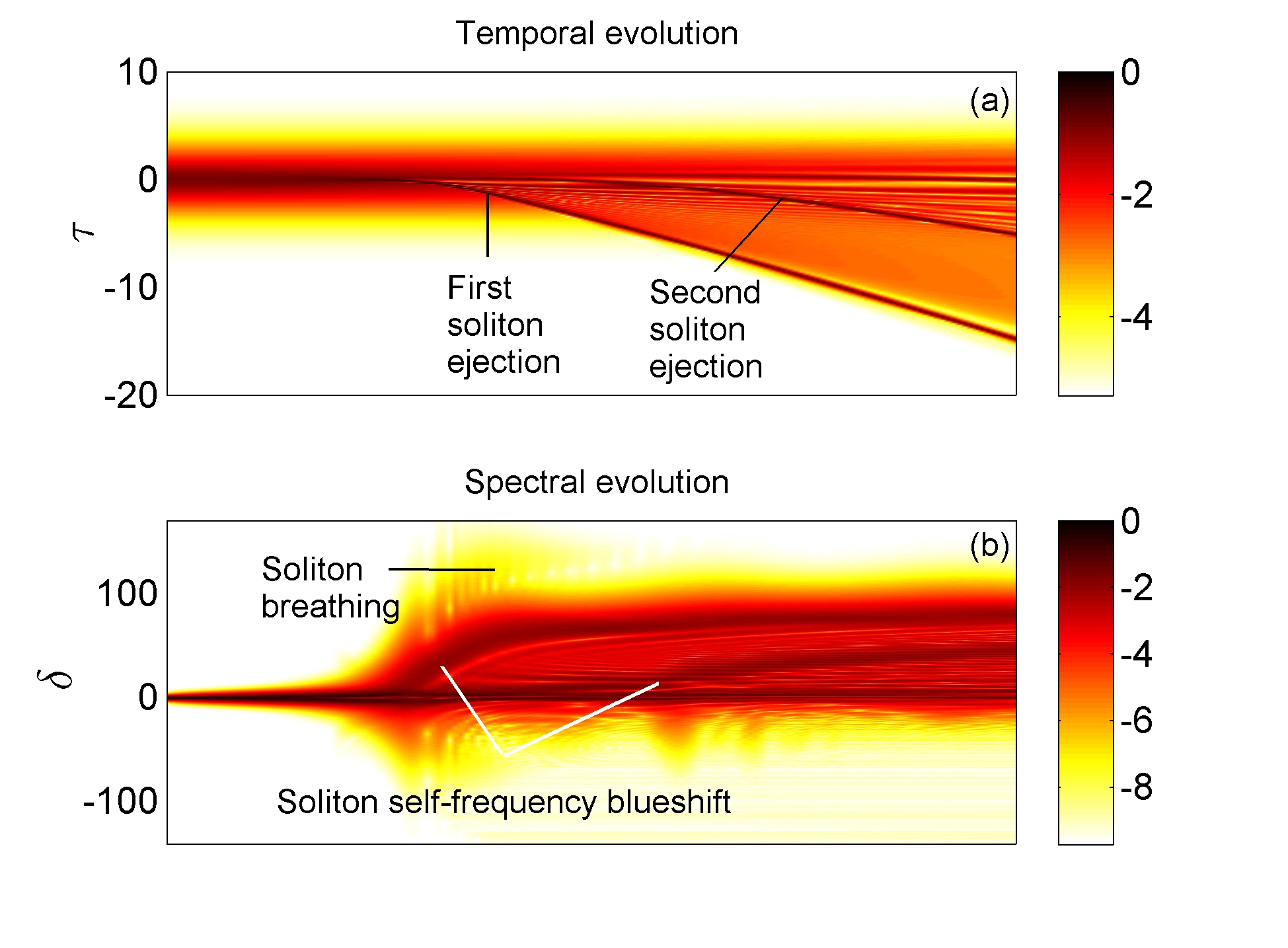}\vspace{-0.65cm}
\includegraphics[width=8.6cm]{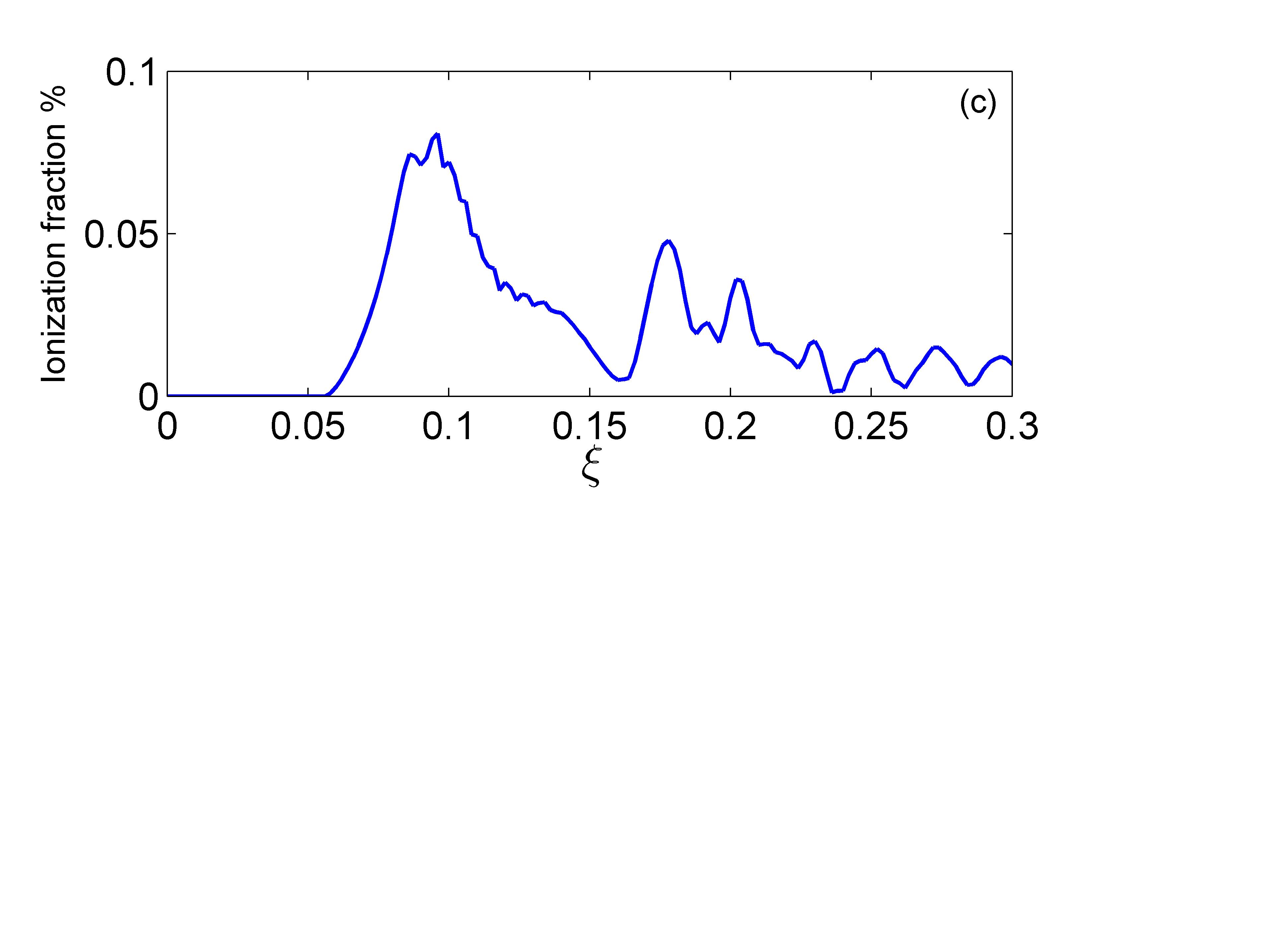}\vspace{-3 cm}
\caption{(Color online). Temporal (a) and spectral (b) evolution of an energetic pulse propagates in an Ar-filled HC-PCF. The temporal profile of the input pulse is $ N\,\mathrm{sech}\,\tau $, with $ N=8 $. The gas pressure is 5 bar. The panels show the ejection of two solitons that continue blue-shifting until the ionization loss suppresses their intensities below the threshold value. Contour plots in this paper are given in a logarithmic scale. (c) Spatial dependence of the ionization fraction along the fiber.
\label{fig4}}
\end{figure}

In this section, we will elaborate our results in \cite{Saleh11} concerning soliton dynamics in an ionized gas. Equations (\ref{eq3},\ref{eq4}) can be numerically solved by using the split-step Fourier method \cite{Agrawal07} in order to fully describe the propagation of pulses in the presence of plasma. In this section, higher-order dispersion coefficients are assumed to be negligible, i.e., $\beta_{m>2}=0  $. The temporal and spectral evolution of a higher-order sech-pulse, with an initial intensity less than the threshold value, are depicted in the panels (a,b) of Fig. \ref{fig4}, respectively. Panel (c) shows the variation of the ionization fraction along the fiber. An optical pulse pumped in the deep anomalous-dispersion regime of the fiber undergoes self-compression. When the pulse intensity exceeds the threshold value, a certain amount of plasma is generated due to gas ionization, and a fundamental soliton is ejected from the input pulse. The soliton central frequency continues to shift towards the blue due to the energy received from the generated plasma. However, due to the concurrent ionization loss, the soliton intensity gradually decreases until it reaches the threshold, thence the pulse blue-shift is ceased. A second ionization event accompanied by a second-soliton emission can take place by further self-compression of the input pulse based on its initial intensity. 

\begin{figure}
\includegraphics[width=8.6cm]{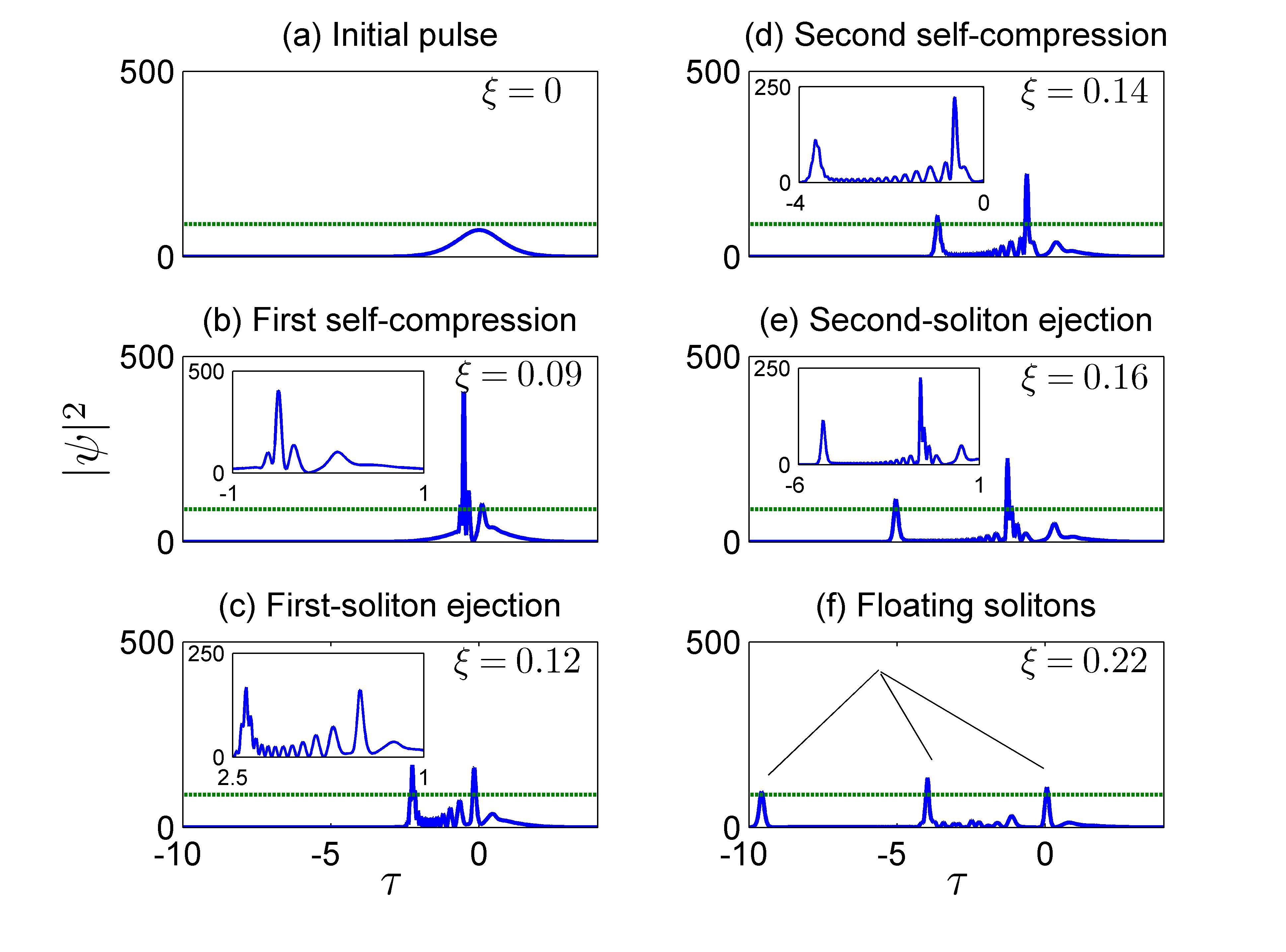}
\caption{(Color online). Intensity profile of a pulse  in the time domain at different positions, $ \xi $, inside an Ar-filled HC-PCF. The dashed red line represents the threshold intensity. The simulation parameters are similar to Fig. \ref{fig4}. Each panel is titled by its main feature. Insets are enclosed in panels for better view and more details. 
\label{fig8}}
\end{figure}

A clear representation for the pulse dynamics in plasma is shown in Fig. \ref{fig8}, where the temporal profile of the pulse intensity $ |\psi|^{2} $ is plotted at selected positions inside the fiber. The simulation parameters are similar to Fig. \ref{fig4}. Initially, $ |\psi(\tau)|^{2} $ is insufficient for plasma ionization. Due to soliton-breathing, the pulse passes through a self-compression stage which strongly enhances its maximum intensity \cite{Tran09}. The amount of optical energy above the threshold intensity $ |\psi|^{2}_{\mathrm{th}} $ contributes to plasma formation that emits a blue-shifted soliton. However due to the ionization-induced loss, the soliton amplitude is attenuated to the regime where $ |\psi(\tau)|^{2} \gtrapprox |\psi|^{2}_{\mathrm{th}} $.  Such pulses are dubbed {\em floating solitons}, since they can propagate for considerably long distance with minimal attenuation and limited blue-shift. Based on the initial input-pulse intensity, other solitons can also be emitted due to further self-compression. At the end, a train of floating solitons are generated. Indeed, Fig. \ref{fig8} clearly shows the formation of a series of floating solitons, that exist and propagate for relatively long distances with their maximum amplitude just above the ionization threshold. These objects are completely unknown in conventional solid-core fiber optics. The fact that these results have been confirmed in concurrent experiments \cite{Hoelzer11b} is a very convincing proof of the validity of our master equations (\ref{eq3},\ref{eq4}).

We have also found an interesting non-local interaction between successive solitons due to the non-vanishing electron density tail, when their temporal separation is shorter than the recombination time. Due to this interaction, a leading soliton can slow down the acceleration of a trailing soliton by an exponential factor. In the frequency domain, the leading soliton suppresses exponentially the blue-shift of the trailing soliton. The reason is that the ionization field $\phi(\tau)$, created by the first soliton, decays at a relatively slow rate. This establishes a unique `non-local' interaction between this soliton and other temporally distant solitons. 

\begin{figure}
\includegraphics[width=8.6cm]{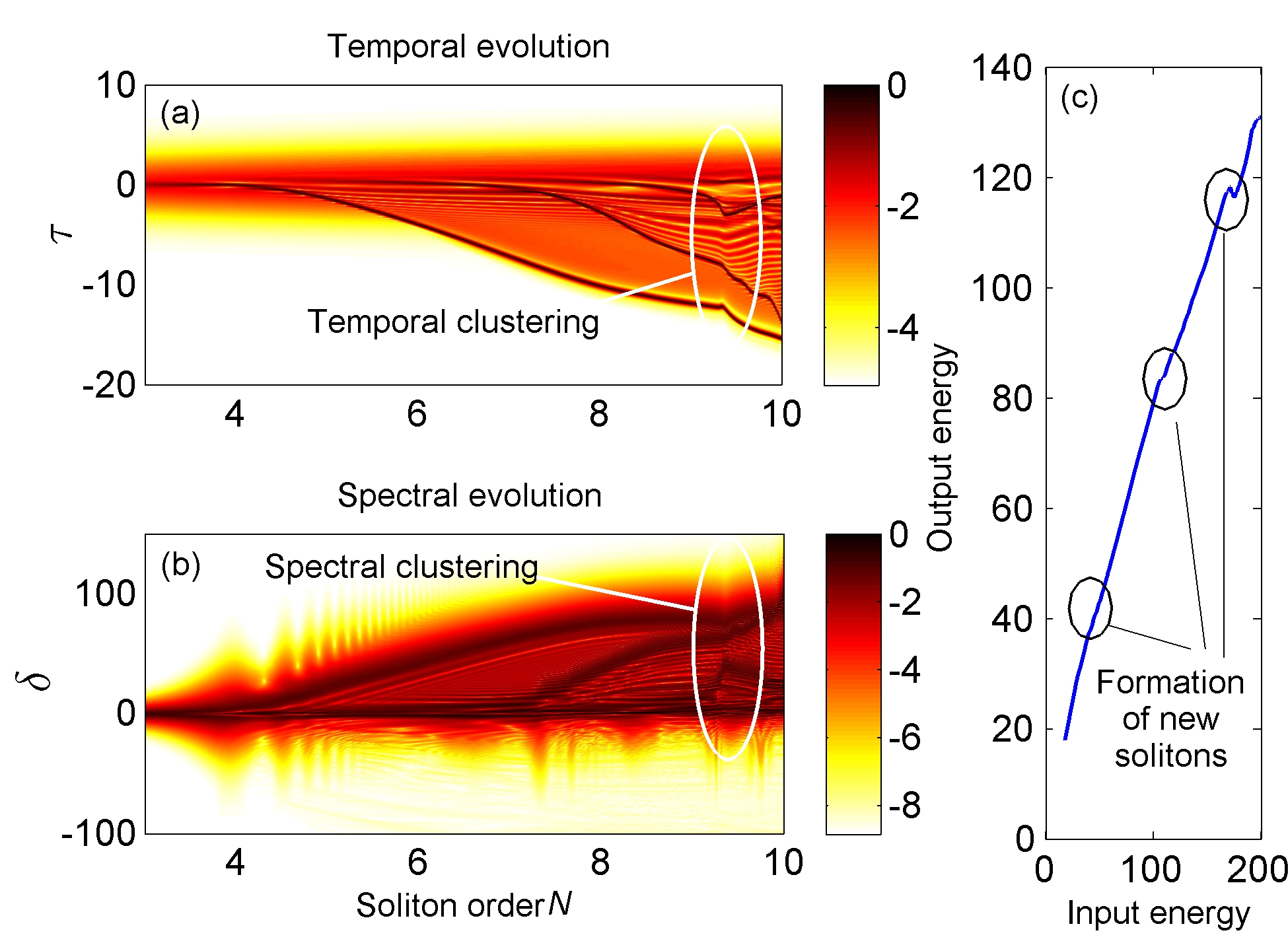}
\caption{(Color online). Dependence of the temporal (a) and spectral (b) outputs of an energetic pulse $ N\,\mathrm{sech}\,\tau $ on the soliton order $ N $. The fiber is an Ar-filled HC-PCF with length $\xi= 1/4 $. The gas pressure is 5 bar. Temporal and spectral clustering occur at $ N=9.2 $ due to the long-range non-local soliton interactions. (c) The output energy versus the input energy.
\label{fig5}}
\end{figure}

These unprecedented dynamics are featured in Fig. \ref{fig5} that show the temporal and spectral dependence on the soliton order $ N $ assuming that the input pulse is $ N\,\mathrm{sech}\,\tau $. A clustering between two or more distant solitons in both temporal and spectral domains is also observed at some `magic' input energy as a result of the interplay between the non-local interaction, ionization loss and ionization threshold. The scenario is as follows: As long as the intensity of the first-emitted soliton, $ I_{1} $, is above the threshold intensity $ I_{\mathrm{th}} $, it prevents the ejection of a second soliton due to the strong effect of the non-local interaction. As soon as, $ I_{1} $ approaches $ I_{\mathrm{th}} $ by virtue of the ionization loss, the first soliton becomes a floating soliton. Hence, the blue-shift and acceleration of this soliton are reduced significantly. Simultaneously, the second soliton is emitted and it can recover its expected acceleration and blue-shift due to the nearly disappearance of the first-soliton non-local force. This allows the second soliton to catch and cluster with the first soliton. In addition, the spectrum of the two solitons start to overlap and form spectral clustering. Similarly, the intensity of the second soliton approaches $ I_{\mathrm{th}} $ due to the ionization loss, allowing a third soliton to cluster with the other two solitons. When the first two solitons are very close to each other, they push back the third soliton due to their \textit{combined} non-local force. In fact, the dynamics after the clustering becomes too complicated to be interpreted in simple terms. Figure \ref{fig5}(c) shows the output energy versus the input energy. The linear dependence of the output energy is slightly broken at the points, which correspond to the ejection of a new soliton. Our results show an excellent qualitative agreement with other non-analytical numerical techniques \cite{Chang11}.

\begin{figure}
\includegraphics[width=8.6cm]{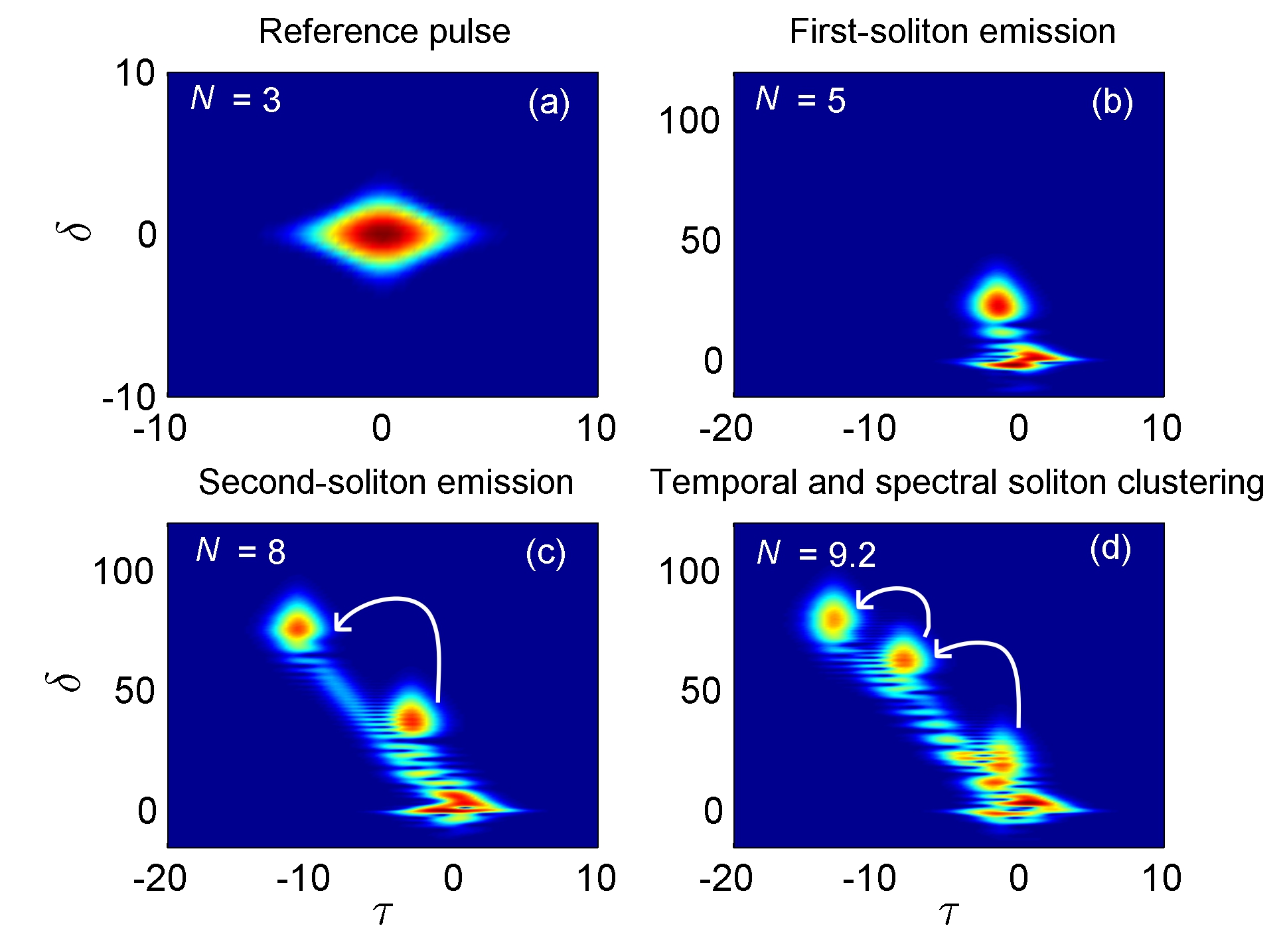}
\caption{(Color online). XFROG spectrograms for pulses with selected soliton order $ N $ --- in an increasing order. The simulation parameters are similar to Fig. \ref{fig5}. (a) $ N=3,\, \xi=0 $. (b) $ N=5,\,  \xi= 1/4$. (c) $ N=8,\,  \xi= 1/4 $. (d) $ N=9.2,\,  \xi= 1/4 $. Each panel is titled by its main feature. White arrows show the movement of the solitons. \label{fig9}}
\end{figure}

XFROG spectrograms for pulses with initial temporal profile $ N\,\mathrm{sech}\,\tau $ and different initial intensities are depicted in the panels of Fig. \ref{fig9}, where (a) represents the reference pulse; (b) and (c) shows the emission of the first and second solitons, respectively; and (d) depicts the temporal and spectral clustering of the first two solitons and the emission of a third soliton. 

\section{Generation of dispersive waves}
\begin{figure}
\includegraphics[width=8.6cm]{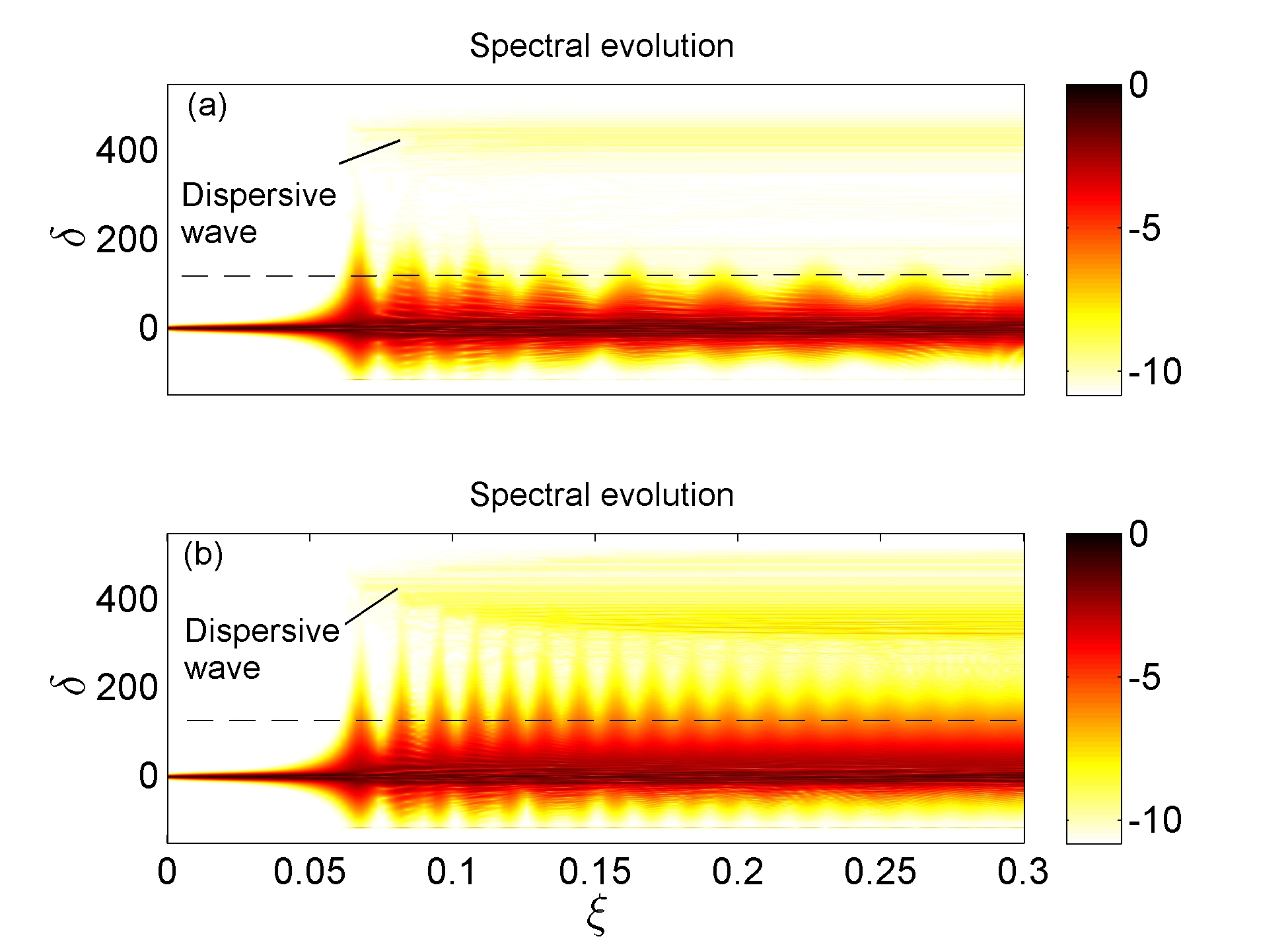}
\caption{(Color online). Spectral evolution of an energetic pulse $ N\,\mathrm{sech}\,\tau $, with $ N=10 $, under the influence of higher-order dispersion coefficients with: (a) switching off the ionization process. (b) switching on the ionization process. The optical pulse central wavelength is $ 0.8\,\mu $m and the gas pressure is 2 bar. The dotted line represents the ZDW.
\label{fig6}}
\end{figure}

The contribution of the higher-order dispersion coefficients $\beta_{m>2}$ to the pulse dynamics starts to play a significant role as pulse central wavelength moves towards the zero dispersion wavelength (ZDW), where $\beta_{2}(\omega_{0})\approx 0$. In fact, higher-order dispersion coefficients may lead to a phase matching condition between two different waves --- an optical pulse, with a central wavelength lies in the anomalous-dispersion regime and is close to the ZDW, and a dispersive resonance wave in the normal-dispersion regime \cite{Husakou01,Cristiani04}. Recently, this fact was implemented to achieve a coherent deep-UV laser source by using an Ar-filled HC-PCF \cite{Joly11}. Joly \textit{et al.} have obtained 8 \% conversion efficiency from IR to deep-UV, where the output can be tuned easily via the pulse energy and gas pressure. The pulse power level was kept below the ionization threshold to avoid any ionization-induced loss. However, we find an enhancement in the dispersive-wave radiation via the photoionization process by more than an order of magnitude due to the ionization-induced blue-shift. As the optical pulse moves towards the blue or the ZDW, new dispersive waves are generated. The conversion to the normal-dispersion regime is ceased when the optical intensity  reaches the threshold value by the ionization loss.  Unlike the Raman process, where the pulse central wavelength is shifted away from the ZDW during pulse propagation results in arresting the optical frequency conversion. The effect of the photoionization on enhancing the generation of the dispersive waves is shown in Fig. \ref{fig6}. The top (bottom) panel represents the spectral evolution of a higher-order sech-pulse in a gas-filled HC-PCF, where the ionization is switched off (on).

\begin{figure}
\includegraphics[width=8.6cm]{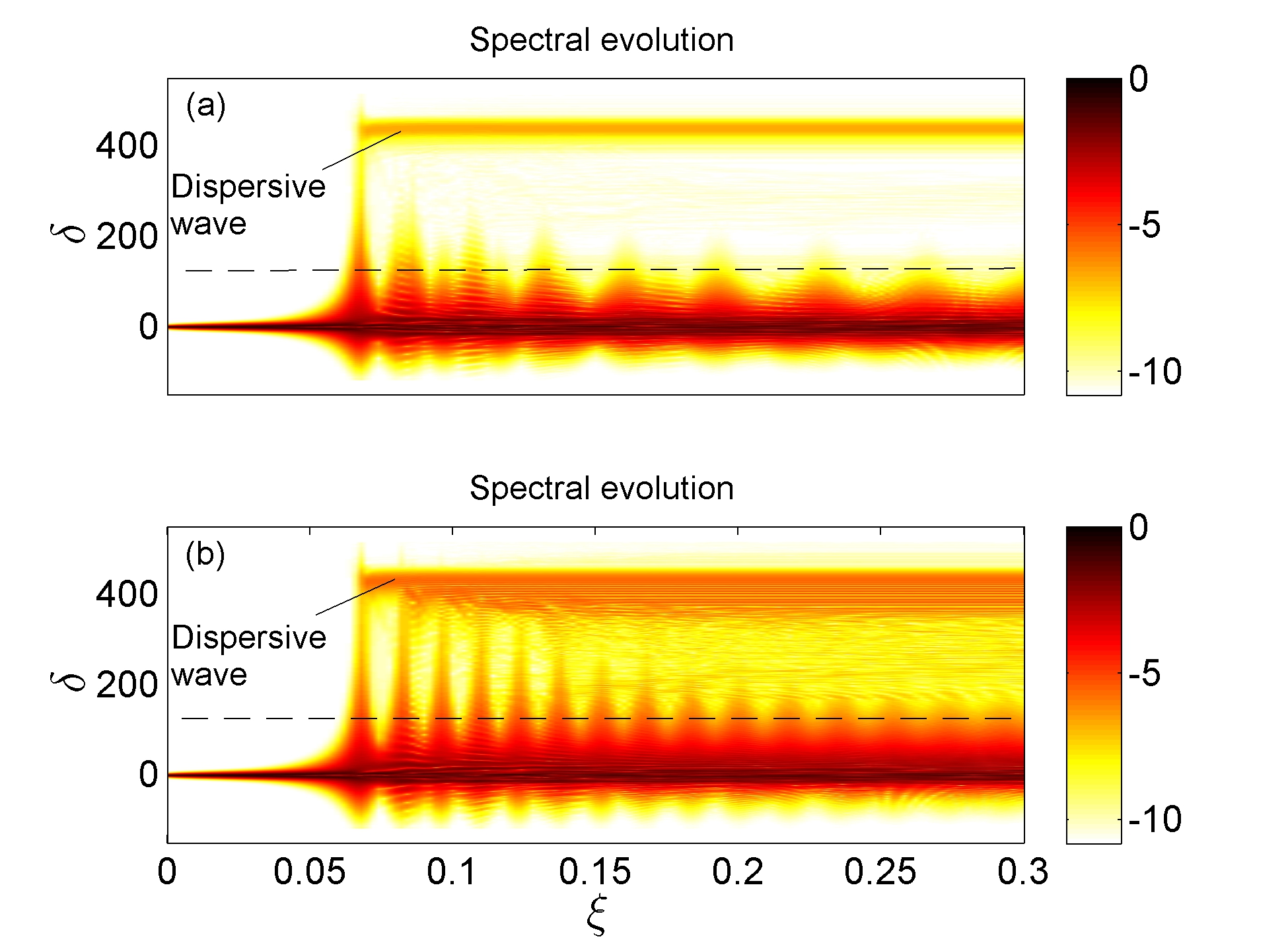}
\caption{(Color online). Spectral evolution of an energetic pulse under the influence of higher-order dispersion coefficients and the shock term with: (a) switching off the ionization process. (b) switching on the ionization process. The simulation parameters are similar to Fig. \ref{fig6} except that the shock operator $ \tau_{\mathrm{shock}} $ is included. The dotted line represents the ZDW.
\label{fig7}}
\end{figure}

The dispersion of Kerr nonlinearity is usually associated with effects such as self-steepening and optical shock formation \cite{Agrawal07}. Due to the absence of Raman effect in kagom\'{e}-style HC-PCFs filled with noble gases, and the weakness of the group velocity dispersion (GVD) in such fibers, the role of the Kerr nonlinearity dispersion assumes an unusual importance. This effect can be studied by using Eqs. (\ref{eq1},\ref{eq2}), where $ \gamma_{\mathrm{K}} $ is replaced by $ \gamma_{\mathrm{K}}\,\left( 1+i\tau_{\mathrm{shock}}\,\partial_{t}\right) $, and $ \tau_{\mathrm{shock}}=1/\omega_{0} $ is the shock time \cite{Agrawal07}. The importance of the shock term is depicted in Fig. \ref{fig7}, where the simulation parameters are similar to Fig. \ref{fig6} except that $ \tau_{\mathrm{shock}} $ is included. Involving the shock term in the pulse dynamics increases the conversion efficiency to the normal-dispersion regime from $ 3\times 10^{-3} \% $ to $ 5\times 10^{-2} \% $ at the end of the fiber due to the spectral asymmetry of the pulse. 

\section{Conclusions}
We have presented a detailed model based on the evolution of the pulse complex envelope to study pulse propagation in gas-filled HC-PCFs under the influence of the nonlinear photoionization-induced effects. By applying perturbation theory, we show that the photoionization process represents the exact counterpart of the Raman self-frequency red-shift of solitons when their intensities are slightly above the threshold intensity. Expressions of the soliton self-frequency blue-shift are derived in the presence and absence of the photoionization threshold. Moreover, the influence of the ionization loss on the soliton amplitude and the frequency shift is studied. Using the Gagnon-B\'{e}langer gauge transformation, stationary negative-slope two-peak `inverted-gravity-like' solitary solutions are obtained for pulses propagate in HC-PCFs filled by Raman-inactive gases. However, positive- or negative-slope two-peak solitary solutions can also be attained in the presence of a tunable Raman-active gas. The pulse dynamics, obtained by using the split-step Fourier method, shows the soliton emission, break-up and blue-shift. Furthermore, we find unconventional long-range non-local interactions between successive solitons due to the non-vanishing electron density tail. The interplay between this unprecedented interaction, the ionization-induced loss and the ionization threshold yields a spectral and temporal clustering between distant solitons. Finally, the effects of higher-order dispersion coefficients and shock term are exploited in generating a strong dispersive-wave radiation in the normal-dispersion regime of the fiber, assisted by the absence of the Raman effect in noble gases.

\acknowledgements 
We would like to thank John Travers, Philip H\"olzer, Wonkeun Chang, Nicolas Joly and Philip St.J. Russell for useful discussions. This research is funded by the German Max Planck Society for the Advancement of Science (MPG).

\bibliographystyle{apsrev4-1}	
%

\end{document}